\documentclass[twocolumn]{revtex4-1}

\usepackage{amssymb,amsfonts,amsmath}
\usepackage{graphicx, graphics}

\begin{document}

\title{Collective sedimentation of colloids in grafted polymer layers} 

\author{Tine Curk$^1$} 
\author{Francisco J. Martinez-Veracoechea$^1$} 
\author{Daan Frenkel$^1$} 
\author{Jure Dobnikar$^{1,2}$}
\affiliation{$^1$Department of Chemistry,
  University of Cambridge, Lensfield Road, CB2 1EW, Cambridge, UK\\
  $^2$Department for theoretical physics, Jo\v zef Stefan Institute, Jamova 39, 1000 Ljubljana, Slovenia} 

\begin{abstract}
We present Monte Carlo simulations of colloidal particles pulled into
grafted polymer layers by external fields. The insertion free energy
of a single colloid into the polymer layer is qualitatively different
for surfaces with an ordered and a disordered distribution of grafting
points. Moreover, the tendency of colloidal particles to traverse the
grafting layer is strongly size dependent. In dense colloidal
suspensions, under the influence of sufficiently strong external
force, a collective instability allows the colloids to penetrate and
form internally ordered, columnar structures spanning the polymer
layer. Depending on the conditions, these colloidal clusters may be
isolated or laterally percolating. The morphology of the observed
patterns can be controlled by the external fields, which opens up
new routes for the design of thin structured films.
\end{abstract}

\keywords{Keywords.}

\maketitle

Grafted polymer layers can prevent the deposition of
colloidal particles on solid surfaces, which is exploited in various
applications like colloidal stabilisation, anti-fouling surfaces
\cite{AntiFouling} and in biomedicine~\cite{BioTech,BioNanoApp}.
Biological surfaces such as the lining of the intestine or the blood
vessels walls are coated with polymers that inhibit the absorption of
too large particles.  Moreover, membrane-attached polymers
over-expressed in cancer cells present a barrier for conventional drug
delivery processes~\cite{cancer}. In several other applications, the
insertion of colloids or nano-particles into polymer layers is
exploited in order to fabricate functional responsive
materials~\cite{Applications_Responsive}: plasmonic
nanostructures~\cite{TunablePlasmonicNano} convert chemo-physical
forces into optical signals; ultrafast switches~\cite{UltraSwitch},
organic memory devices~\cite{ElectricalSwitches} and advanced
photovoltaics~\cite{PolySolar, PhotoVoltaics} are based on
incorporating metallic particles into polymers.

The physical properties of grafted polymers layers are governed by the
configurational entropy of tethered polymers and have been studied
extensively for different grafting and solvent
conditions~\cite{deGennesBook,MilnerWittenCates,Zhao,Ruhe}. The
penetration of particles into polymer layers has been recently
addressed in theoretical and numerical
studies~\cite{Ermilov,Binder2}. The insertion free energy of an
isolated protein~\cite{ProteinAdsorption,Ivan_Mark} or colloid
\cite{ChenChen,Kim2008} has been explored for varying grafting
densitiies, particle sizes and shapes~\cite{Kim2006}. Colloids soluble
in polymers penetrate up to a depth determined by their
size~\cite{KimPRL2002}, while insoluble particles form aggregates,
which are expelled from the brush after reaching a critical
size~\cite{Guskova}. The polymer-mediated colloidal interactions are
typically of the order of $k_BT$~\cite{Kim2006,pair} (i.e., comparable
to the entropic terms). The colloid-polymer mixtures are thus
inherently less ordered than molecular systems. External forces could
potentially be used to re-gain the order: due to the constrained
translational degrees of freedom of the colloids pulled into the
layers, the weak polymer-induced interactions could steer the assembly
into ordered patterns. However, little is known about the collective
particle sedimentation in a brush under the influence of a constant
force.

\begin{figure} 
  \centering
\centerline{ \includegraphics[width=8.5cm]{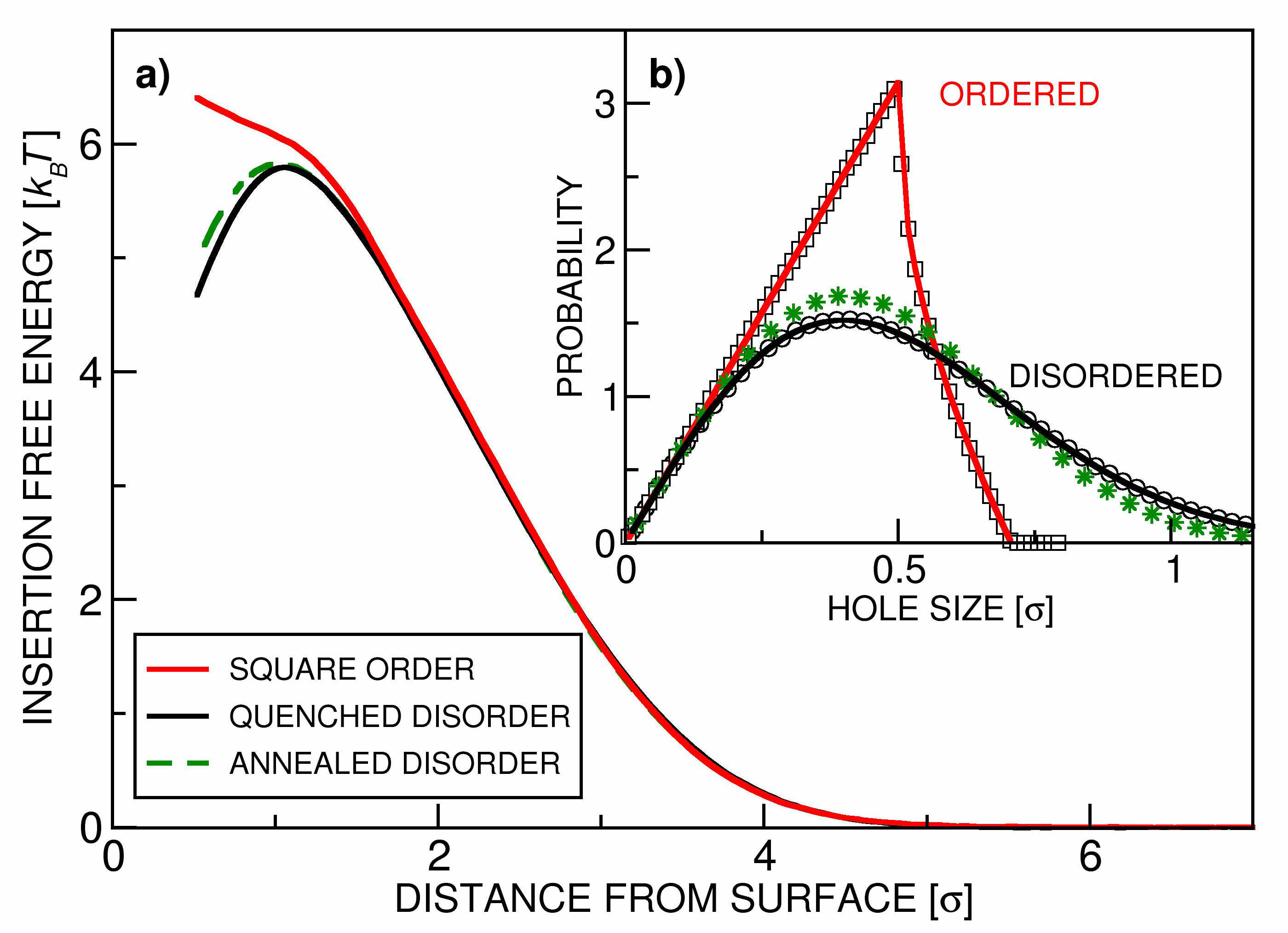} }
 \caption{Isolated colloids in polymer layers with grafting density
   $\rho=1.0$ and number of blobs per chain $l_p=20$. {\bf (a)} The
   insertion free energy profile $F_p(z)$ for three realizations of
   the surface grafting: quenched disorder (black solid line), square
   crystaline order (red solid line), and annealed disorder (green
   dashed line). {\bf (b)} The distribution of (2D) hole sizes for the
   three scenario. The solid lines represent the analytically derived
   expressions (see {\sl Online SI}), symbols are numerically
   determined.}
 \label{Fig1}
\end{figure}
Here we report Monte Carlo simulations of polymer-insoluble colloidal
particles in grafted polymer layers. The system comprises of polymer
chains end-grafted to a flat surface and hard sphere-colloids with
diameter $\sigma$ immersed in a neutral good solvent. The colloids
interact via hard-core and polymer-mediated interactions and are
additionally subject to a constant external force $f_{g}$, whose
significance is measured by the ratio of potential versus thermal
energy: $g'= f_g\sigma/k_BT$. A possible example of the external force
is gravity: $f_g=g (\rho_c-\rho_s)\pi \sigma^3/6$, with
$\rho_c-\rho_s$ the excess colloidal density in the solvent and $g$
the standard gravitational acceleration. For gravity, the ratio
$g'\propto \sigma^4$ strongly depends on the colloidal size: it is
efficient for large-enough (micron-size) colloids and does not affect
nano-particles or polymers in solution. Polymers can thus be treated
as self-avoiding random walks. We follow a coarse-grained
model~\cite{BarbarasBlobs} where the self-avoiding polymer chains are
represented by $l_p$ soft blobs with radius of gyration $r_b$
\cite{Blob_Interactions}. The grafting density at the surface,
$\tilde{\rho}=N_{p}/R_g^{2}$, determines whether the layer is in the
dilute ``mushroom" ($\tilde{\rho} < 1$) or in the dense "brush"
scaling regime ($\tilde{\rho} \gtrsim
3$)~\cite{Binder,TetheredChains}. We studied the transition regime
with the mean spacing between the grafting points roughly similar to
the polymer $R_g$~\cite{Length_Scale}. Expressing the grafting density
as number of anchors per colloidal diameter squared:
$\rho=\tilde{\rho} R_g^2/\sigma^2$, we focused on the regime $0.5 <
\rho < 5$, where the probability of inserting a colloid with
$\sigma\approx R_g$ is non-vanishing.

The free energy profile $F(z)$ of a single colloid is the sum of its
potential energy in the external field and the insertion free energy
$F_p(z)$:
\begin{equation}
\beta F(z) = g' \frac{z}{\sigma} + \beta F_{p}(z) \;.
\label{Eq:Ftot}
\end{equation}
The potential energy term is trivial and decouples from the polymeric
degrees of freedom, therefore the insertion free energy is the key
quantity governing the penetration of colloids into the polymer
layers. We have used the above described coarse-grained model of
polymers and evaluated the insertion free energy $F_p(z)$ by the
Wang-Landau technique~\cite{WangLandau} for a wide range of polymer
grafting densities $\rho$ and number of blobs per chain $l_p$. In
Figure 1{\bf (a)} we compare polymer brushes characterized by ordered
and disordered arrangement of the grafting points: $F_{p}(z)$ is
plotted for layers with identical $\rho$ but different spatial
distribution of the grafting points: ``quenched disorder'' (random
Poissonian process), ``annealed disorder'' (random grafting points
relaxed before the simulation), and ``order'' (grafting points on a
square lattice). The insertion free energy is qualitatively different
in ordered and disordered case: the repulsion is strongest in case of
ordered grafting where the free energy is monotonically decreasing
with the height. In disordered case the free energy features a
barrier. The height and position of the barrier and the depth of the
minimum at the surface can be controlled (see {\sl Online SI}), which
enables reversible adsorption and slow release of the colloids at
disordered surfaces.

The emergence of the free energy barrier (also observed in other
simulations \cite{Ermilov,Ivan_Mark}) is surprising: the mean monomer
concentration profile is equal for ordered and disordered grafting
and, at $\rho=1.0$, it is monotonically decreasing with height (see
{\sl Online SI} and~\cite{MuratGrest1989,Binder2}). The insertion free
energy is obtained by exponential averaging of such profiles and is
determined by both, the mean monomer concentration and its
fluctuations. The fluctuations are characterized by the probability
distribution of holes (regions void of monomers)\cite{Mark-JCP-09}
. In the inset (Figure 1{\bf (b)}) the 2D hole-size distributions at
the bottom surface are plotted: in case of disordered grafting
characterized by the Poissonian statistics, there is an appreciable
probability of finding large holes, while the ordered surfaces are
characterized by small density fluctuations with a cut-off in the hole
size. Close to the grafting surface the monomer fluctuations are
``frozen'' because the anchoring points are immobile, while away from
it they relax: the surface-imposed effects vanish at a height equal to
the characteristic length-scale~\cite{Length_Scale}. The free energy
cost of inserting a particle into a disordered brush therefore
initially grows as a function of height and eventually decreases along
with the average monomer concentration - resulting in the barrier with
a maximum around $z\approx R_g$. The barrier height scales with $\rho$
and with $\sigma^2$; in our system its typical magnitude is several 10
$k_BT$ (see {\sl Online SI}).

The potential energy term (\ref{Eq:Ftot}) significantly alters the
free energy landscape: for large enough $g'$ the unstable local
minimum becomes stable and the colloids are strongly confined to the
surface. Since in the case of gravity $g'$ sensitively depends on the 
colloidal size, this
property can be used for particle-size sorting. In Figure 2 we plot
the total free energy profiles {\bf (a)} for colloids of three
different sizes together with their height probability distribution
{\bf (b)}. Counterintuitively, the smallest colloids gather on top of the brush 
(the gravity is too week to overcome the steric barrier), while
slightly larger colloids are distributed in a bimodal way and even
larger ones are strongly confined to the bottom surface. This
single-colloid picture also governs the density profiles in
semi-dilute systems as seen in the simulation snapshots in Figure
2{\bf c)}).
\begin{figure}
\vspace{0.2cm}
 \centerline{\includegraphics[width=8.5cm]{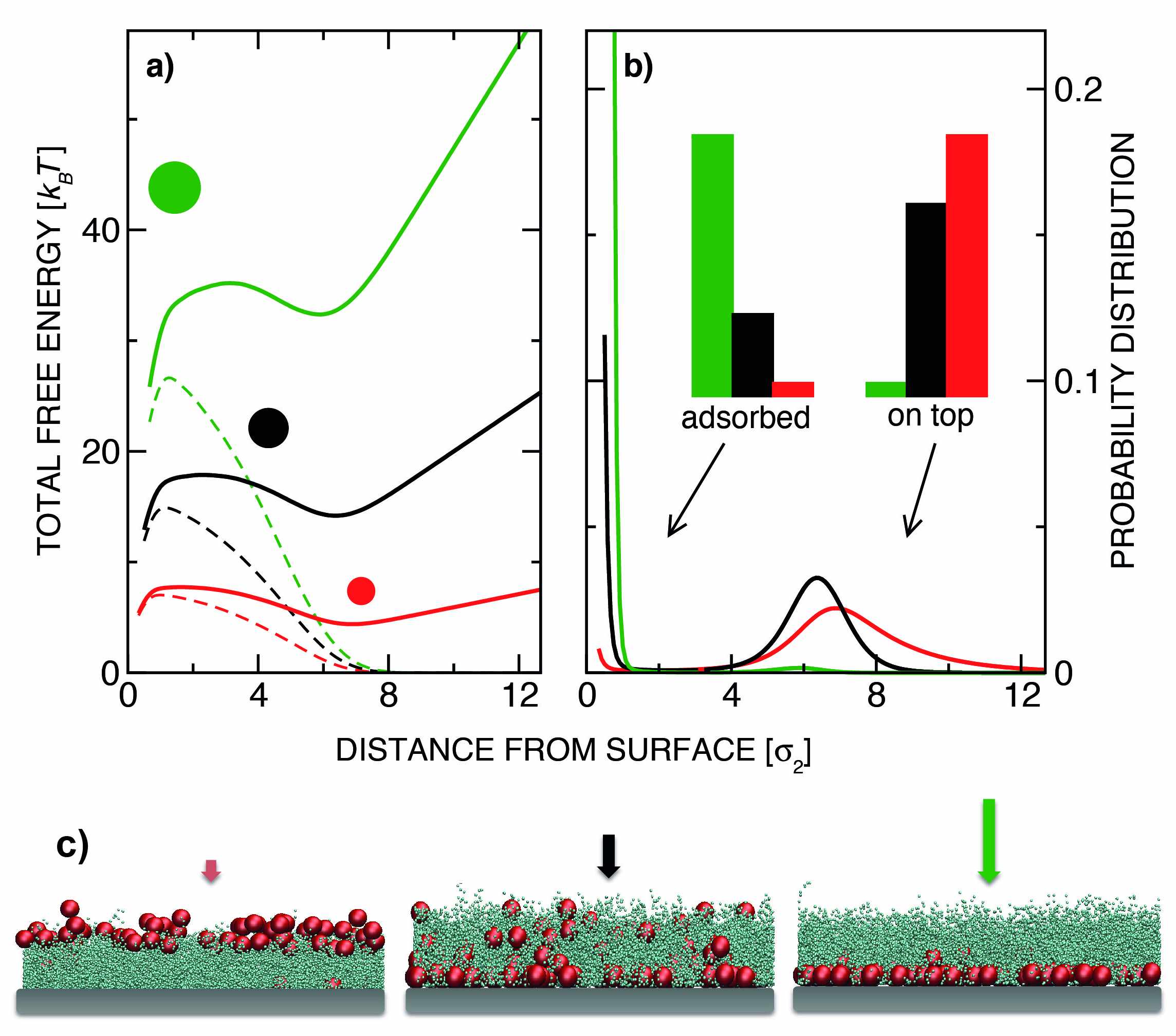} }
\caption{{\bf (a)} Total free energy (\ref{Eq:Ftot}) of sedimenting
  colloidal particles obtained by simulations at $l_{p}=40$ and
  $\rho=1.0$ for silica colloids with three sizes: $\sigma_1 = 0.67
  {\rm \mu m}$ (red line), $\sigma_2 = 1.00 {\rm \mu m}$ (black line),
  and $\sigma_3= 1.33 {\rm \mu m}$ (green line). Respectively,
  $g'_{2}=2.0$ (corresponds to micrometer silica particles in water),
  $g'_1=0.40$ and $g'_3=6.3$.  The dashed lines represent the
  insertion free energy $F_p(z)$.  {\bf (b)} Height distribution
  $P(z)\propto e^{-\beta F(z)}$ for the same system. The histograms
    illustrate the probability to find the colloids adsorbed to the
    surface or above the brush (left set of histograms: $z < R_g$ and
    right set: $z>R_g$).  {\bf (c)} Snapshots from Monte Carlo
    simulations of dilute colloidal suspensions at conditions
    corresponding to {\bf (a)} and {\bf (b)}.}
  \label{Fig2}
\end{figure}
The amount of particles drawn into the polymer layer reflects the
balance between the polymer-induced pressure pushing the particles
away and the external forces pulling them towards the surface. At a
critical grafting density $\rho=\rho_0$ the two effects are balanced:
for $\rho>\rho_0$ no particles can penetrate the polymer
layer~\cite{Rho_0}. For $\rho<\rho_0$ the particles are pulled in the
polymer layer by the excess external pressure - thereby displacing the
polymers and increasing their effective density. When many colloids
are inserted, the polymers mediate effective many-body interactions
among the particles and the insertion free energy is no longer described by the
single-particle term $F_p(z)$. In general it
depends on the positions of all particles. To explore the
multi-particle systems we have performed Grand-canonical Monte Carlo
simulations with colloidal particles coupled to a reservoir with a
fixed chemical potential at various polymer grafting densities, chain
lengths, particle concentrations and external forces.

At large-enough pulling $g'$ the penetrating particles start forming
crystalline clusters at the bottom surface that grow towards the top
of the brush~\cite{kinetics}. Figure 3 shows typical growth scenarios
on disordered and ordered surfaces. In the initial stages there is a
difference: the shape of the clusters above the ordered surfaces is
narrower at the bottom (resembling a table-top) due to the strong
repulsion of colloids from the surface, while in case of disordered
grafting the local free energy minimum enables pyramid-like
structures. In both cases, however, similar final structures emerge:
clusters with a uniform vertical profile spanning the brush from
bottom to top resembling straight cylindrical towers.
\begin{figure}
\centerline{\includegraphics[width=9cm]{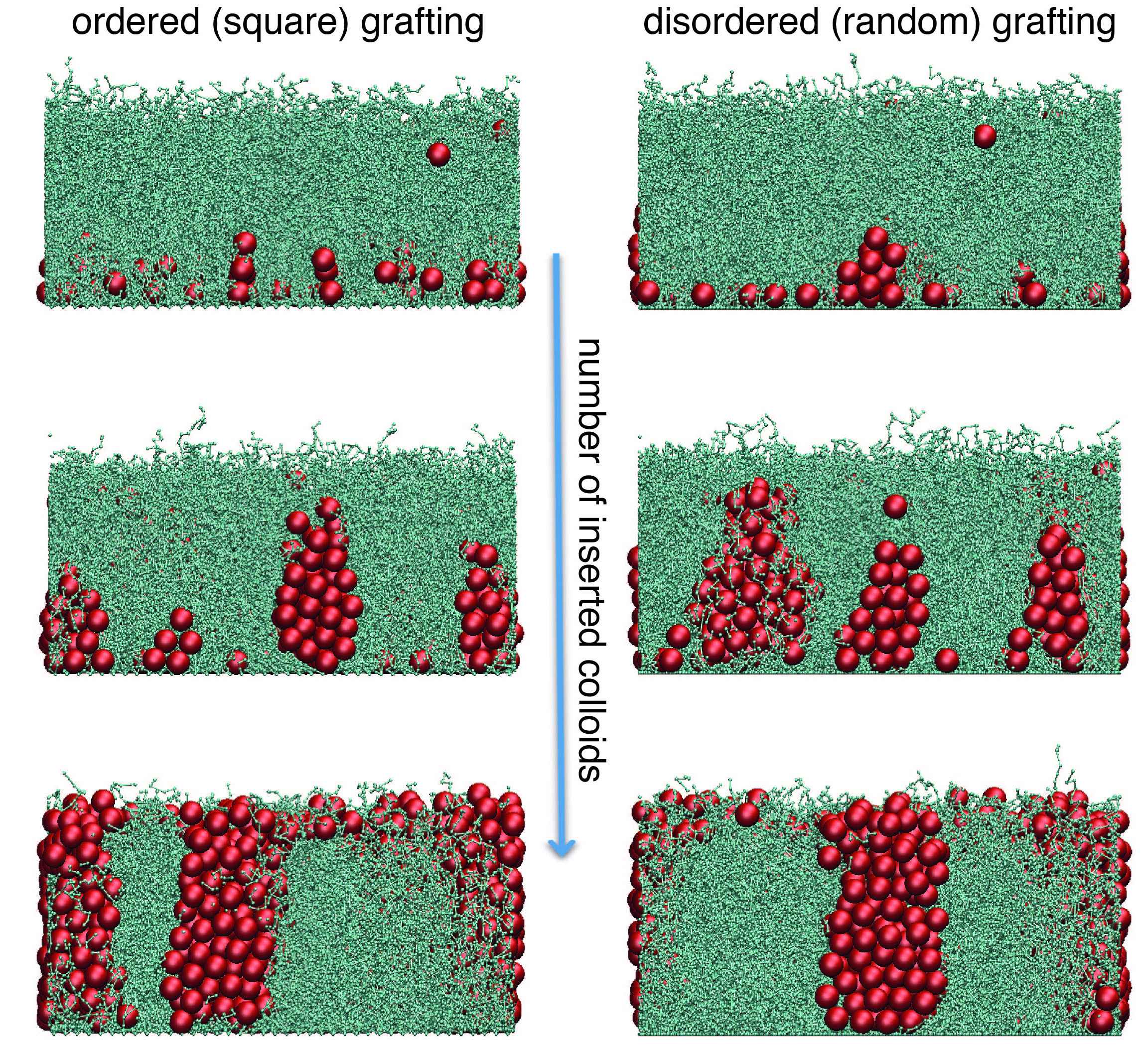} }
\caption{Cluster growth as a function of the number of colloids in the
  system.  Sedimentation sequence on ordered (left-hand side) and
  disordered (right-hand side) surfaces is depicted. Initially,
  ``pyramide'' and ``table-top'' structures are formed on disordered
  and ordered surfaces, respectively. Eventually, in both cases
  ``critical'' clusters with uniform vertical profiles spanning the
  polymer layer are formed. The number of blobs per chain is $l_p=40$,
  the anchoring density $\rho=4.0$, and the effective gravity
  $g'=6.0$.}
  \label{Fig4}
\end{figure}
The maximum load of colloids in the polymer layer is achieved when the
effective polymer density in the space around the colloids equals
$\rho_0$. Consequently, the thickness of a fully loaded layer $\propto
\rho_0^{0.35}$~\cite{deGennesBook} does not depend on the polymer
grafting density and can be controlled by inserting particles into the
polymer layer (experimentally, an excessive amount of colloids could
be added and the ones above the brush could be simply
washed-off). Given the uniform vertical profiles observed in the
colloidal clusters (Figure 3), the particle load is directly
correlated to the colloidal surface coverage $\eta$ of the two-dimensional
horizontal cross-sections (top-view snapshots in Figure 4):
$\eta\equiv S_{col}/S = 1- \frac{\rho}{\rho_{0}} $, where $S_{col}/S$
is the fraction of the area on the snapshot covered by the (red)
colloids. 

In Figure 4 we present the phase diagram as a function of the grafting
density $\rho$ and the critical density $\rho_0 \propto g'$.  We have
investigated both, ordered and disordered surfaces and observed no
important differences in the lateral morphology. Different structures
observed in the simulations are depicted by symbols and their phase
boundaries are approximately marked by the solid lines, which are
lines of constant colloidal surface coverage $\eta$. The white lower
region of the diagram is the ``super-critical'' grafting region where
no colloids penetrate the brush and no ordered structures are
formed. The grey-shaded region depicts the regime where the colloids
can penetrate the brush but the effective polymer density is too small
and the entropy prevents particle ordering \cite{Fluid}. 
\begin{figure}
\centerline{\includegraphics[angle=90,width=9cm]{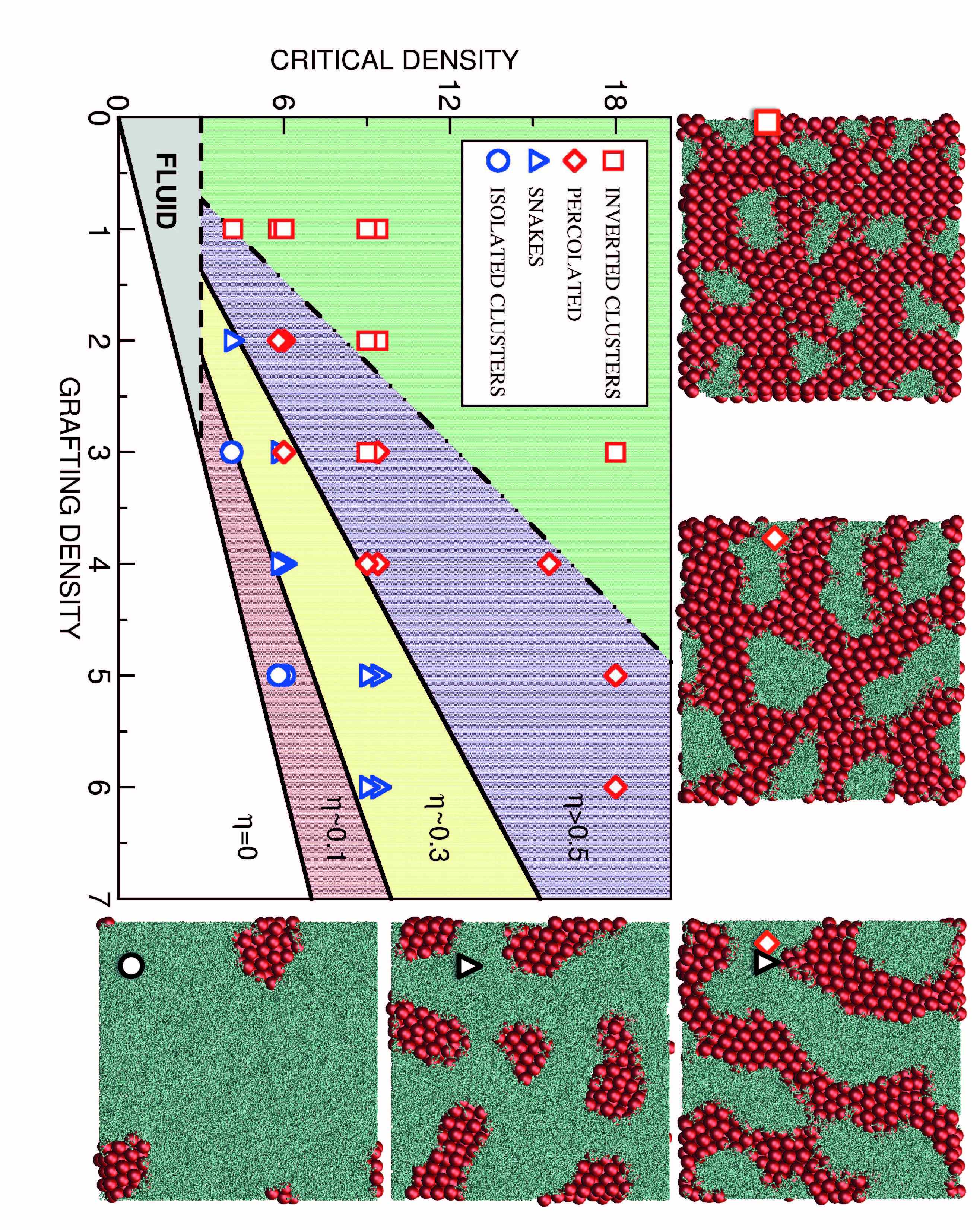} }
\caption{Phase behaviour as a function of the polymer grafting density
  and the external pressure. Various lateral patterns are depicted by
  symbols: isolated cylindrical towers (blue circles), ``snakes''
  (blue triangles), percolating structures (red diamonds) and inverted
  towers (red squares).  The black lines of constant surface coverage
  $\eta$ separating the phases are guides to the eye. Top-view
  snapshots from the simulations displaying the lateral morphology are
  shown at the side and marked by the same symbols. The snapshots are
  taken at an intermediate height within the polymer layer and
  correspond to the value of the critical density $\rho_0=6.0$ with
  $g'=6.0$, $l_p=40$ and the grafting density ranging from $\rho=1.0$
  (inverted towers) to $\rho=5.0$ (isolated towers). The typical
  length-scale and the strong crystalline order are clearly revealed.}
\label{Fig5}
\end{figure}
In order to understand the micro-phase separation leading to these
various patterns we consider two mechanisms: the depletion of polymers
around solid objects that favours macro-phase separation, and an
elastic-like penalty due to the fact that the polymers are grafted
that promotes colloid-polymer mixing. The competition between such
terms, combined with the action of the external forces, results in the
formation of finite and ordered lateral patterns uniquely
characterized by the colloiodal surface coverage: at supercritical
conditions ($\eta\equiv 0$) the colloids are in a liquid state on top
of the brush; at small $\eta$ isolated circular clusters emerge. Upon
increasing the colloidal surface coverage snake-like objects,
percolating structures and finally inverted clusters are formed. The
morphologies are quite robust and are characterised by a well-defined
length-scale (the width of the towers, snakes, or percolating
structures). Interestingly, this length-scale does not depend on the
grafting density but scales with $l_p$ (i.e. with the polymer size)
and is roughly equal to the brush height. The stability and properties
of the reported patterns is reproduced by a phenomenological theory
presented in {\em Online SI}.

Our results provide an insight into many-body polymer-induced particle
interactions, which are of key relevance for understanding various
biological processes, and to design novel materials and biomedical
applications. Metal-coated micro-particles pulled into grafted layers
by gravity can form thin metallic films with unique mechanical,
electric conductivity or heat transport properties. If mechanical
stress can be applied to stretch or bend the substrate, the grafting
density and the morphology of the emerging structures can be varied in
a controlled way, which should provide a useful insight for surface
characterization. We furthermore predict that grafted surfaces can be
used as a sorting device for polydisperse particles or particle
aggregates, as well as to control the rates of chemical reactions. It
has been shown \cite{Ivan_Mark} that the polymer brushes can affect 
protein folding processes, which is not surprising in light of
our results. In order to fabricate the predicted multi-particle
structures with nano- instead of micro-particles, ultra centrifuge can
be used instead of gravity. The morphology of the reported structures
is strongly desirable in modern nano-responsive applications like
photovoltaics~\cite{PhotoVoltaics2}, smart glasses~\cite{UltraSwitch}
and miniature sensors~\cite{Oren}.

\begin{acknowledgments}
We gratefully acknowledge enlightening discussions with Oren Scherman,
Bortolo M. Mognetti, Patrick Varilly, Stefano Angioletti-Uberti, and
Mark Miller. This work was supported by the 7th Framework Program of
European Union through grants ARG-ERC-COLSTRUCTION 227758 (D.F.,
F.M.V., J.D.) and ITN-COMPLOIDS 234810 (J.D., T.C.), by the Erasmus
student exchange program (T.C.), and by the Slovenian research agency
through Grant P1-0055 (J.D.).
\end{acknowledgments}


\begin{thebibliography}{10}
\bibitem{Blob_Interactions} The blobs are connected with springs:
  $U_{ch}/k_BT = 0.534 (r/r_b -0.730)^2$ and repel via $U_{bb}/k_BT
  =1.75\, e^{-0.80 (r/r_b)^2}$. They are also repelled from the
  colloids and surfaces via $U_{bc}/k_BT =3.20\,
  e^{-4.17(r/r_b-0.50)}$.
\bibitem{Length_Scale} The relevant length scale in the dilute
  ``mushroom'' regime is the radius of gyration $R_g$ of the
  chains. In the brush regime it is the so-called de Gennes blob (mean
  distance between the grafting points),
  $\xi\propto\rho^{-1/2}$~\cite{deGennesBook}. All our simulations are
  in the intermediate regime where the two length-scales are similar.
\bibitem{Rho_0} The value of $\rho_0$ at a given externally applied
  force depends on the the solvent properties and on details of the
  colloidal and polymeric interactions. In our model the former are
  hard-core and the latter were chosen to describe the self-avoiding
  walk.
\bibitem{kinetics} The colloidal insertion and deletion has been
  attempted with various rates. The attempted insertions were at
  random positions. However, most accepted insertions were on top of
  the brush and the colloids later diffused into the layer. At large
  $\rho$ the system is frequently trapped in metastable states and the
  observed structures depend on the rate of the attempted
  insertions. At high rate, colloids form a layer on top compressing
  the polymers below them. If the effective density of polymers is
  thereby increased above critical $\rho_0$, no colloids can
  penetrate. For slow-enough insertion, the colloids
  inserted on top have enough time to explore the free energy
  landscape and the system can reach the thermal equilibrium.
\bibitem{Fluid} The extent of the fluid region depends on the physical
  parameters, especially on the colloidal excess density and
  size. Presented in Figure 4 is the case of $\sigma=1 {\rm \mu m}$
  silica colloids in water; for smaller colloids the boundary of the
  fluid region would would shift upwards.
\end{thebibliography}
\end{document}